\documentclass[10pt,leqno]{amsart}
\usepackage{graphicx}
\usepackage{indentfirst}

\usepackage{amssymb,amsthm,amsmath}
\usepackage{cite}
\usepackage{xcolor,paralist,hyperref,titlesec,fancyhdr,etoolbox}
\usepackage[utf8]{inputenc}
\usepackage[english]{babel}
\usepackage[perpage]{footmisc}
\usepackage{subcaption}

\titleformat{\section}[display]{\normalfont\huge\bfseries\centering}{}{10pt}{\Large}
\titlespacing*{\section}{0pt}{0ex}{0ex}

\hypersetup{ colorlinks=true, linkcolor=black, filecolor=black, urlcolor=black }

\begin{document}

\bibliographystyle{unsrt}

\title{Magnetically Induced Current Density from Numerical Positional Derivatives of Nucleus Independent Chemical Shifts} 

\author{Raphael J.F. Berger\textsuperscript{1} \and Maria Dimitrova\textsuperscript{2}}

\date{\today}

\maketitle
\begin{center}
\textsuperscript{1}Fachbereich f\"ur Chemie und Physik der Materialien, Paris-Lodron Universit\"at Salzburg, Jakob-Haringerstr. 2a, A-5020 Salzburg, \"Osterreich.\\
\textsuperscript{2}Department of Chemistry, Faculty of Science, FI-00014 University of Helsinki, P.O. Box 55, A. I. Virtasen aukio 1, Finland.\\
\end{center}
\begin{center}
\textsuperscript{1}\texttt{raphael.berger@plus.ac.at}, \textsuperscript{2}\texttt{maria.dimitrova@helsinki.fi}
\end{center}


\begin{abstract}
  Instead of computing magneticallly induced (MI) current densities (CD)
  via the wave function and their quatum mechanical definition one can also use
  the differential form of the Ampère-Maxwell law to obtain them from spatial derivatives
  of the induced magnetic field. In magnetic molecular response calculations,
  the latter can be done by numerical derivativation of the so called
  ``nucleus-independent chemical shifts'' (NICS) which are avaialable
  to many standard quantum chemical programs. The resulting numerical MICD
  data is in contrast to other numerically obtained MICDs computed via the wave function
  route, virtually divergence-free.
\end{abstract} 
\maketitle

\bigskip

\section{Introduction}

At the conclusion of the fourth and final paper in Schrödinger's seminal series "Quantisierung als Eigenwertproblem" \cite{Schroedinger1926-IV}, he introduced a vector quantity that is like the density function bilinear in the wave function and its complex conjugate. He interpreted this quantity as the current density ({\em Stromdichte}) associated with the probability density ({\em Gewichtsfunktion}) in configuration space, writing: “welcher offenbar als die Stromdichte der Gewichtsfunktion im Konfigurationsraum zu interpretieren ist.” Schrödinger further concluded that this current density vanishes for nondegenerate energy eigenstates, leading to his strikingly simple explanation of the radiationlessness of atomic ground states: “Damit findet die Strahlungslosigkeit des Normalzustandes allerdings eine verblüffend einfache Lösung.”

The current density, which we abbreviate as CD or $\mathbf{J}$, has been of central importance to quantum theory from its earliest days to the present. In modern notation, and for the one-particle case it is defined as

\begin{equation}
  \mathbf{J} := \Re{\{\psi^* \boldsymbol{\pi} \psi\}}
\end{equation}

where $\psi$ is the quantum mechanical wave function describing the state of the particle and $\boldsymbol{\pi}$ is the canonical momentum operator\footnote{which itself corresponds to the conjugate of a spatial degree of freedom of the Langrangian} 

Today, for example, the CD plays a critical role in Theoretical Chemistry, as it encodes the complete information on molecular magnetic response.\cite{RingCurrents,Berger1,Berger2} All physical magnetic properties, such as magnetic susceptibilities and shielding constants, can be derived directly from the CD.

In quantum chemistry, the computation of $\mathbf{J}$ has traditionally relied on Schrödinger's original defining equation, using the wave function as the starting point. Virtually all quantum chemical codes and programs to date employ this approach.\cite{Berger1}

We propose an alternative strategy, inspired by Hirschfelder's notion that his so-called "subobservables"\cite{Hirschfelder} can be treated analogously to classical quantities\cite{RingCurrents}. This perspective in conjunction with the electrodynamic field equations offers a fresh framework for deriving $\mathbf{J}$ potentially opening up new computational and conceptual pathways. 

\section{Results and Discussion}

Electrons in a molecule respond to a weak external magnetic field $\mathbf{B}_{ext}$ by inducing a secondary magnetic field $\mathbf{B}_{ind}$, such that in every point $\mathbf{r}$ in space a total magnetic field $\mathbf{B}_{tot}$ results. These fields are related via the so called ``chemical shift tensor'' $\mathbf{\sigma}(\mathbf{r})$, describing the magnetic response of the molecule

\begin{align}
  \mathbf{B}_{tot}(\mathbf{r}) & = (\mathbf{1}-\boldsymbol{\sigma}(\mathbf{r}))\cdot\mathbf{B}_{ext}(\mathbf{r}) \\
                               & = \mathbf{B}_{ext}(\mathbf{r}) - \underbrace{\boldsymbol{\sigma}(\mathbf{r})\cdot\mathbf{B}_{ext}(\mathbf{r})}_{= \mathbf{B}_{ind}(\mathbf{r})}.
\end{align}

The chemical shift tensor can be directly related to the ``nucleus-independent chemical shift'' tensor or NICS \cite{NICS1,NICS2} via

\begin{align}
  \boldsymbol{\sigma}(\mathbf{r}) & = - \begin{pmatrix} \text{NICS}_{xx}(\mathbf{r}) & \text{NICS}_{xy}(\mathbf{r}) & \text{NICS}_{xz}(\mathbf{r}) \\
    \text{NICS}_{yx}(\mathbf{r}) & \text{NICS}_{yy}(\mathbf{r}) & \text{NICS}_{yz}(\mathbf{r}) \\
     \text{NICS}_{zx}(\mathbf{r}) & \text{NICS}_{zy}(\mathbf{r}) & \text{NICS}_{zz}(\mathbf{r})
  \end{pmatrix}
\end{align}
where the second index refers to the external field. In tensor notation this equals to
\begin{align}
   \boldsymbol{\sigma}_{\alpha\beta}(\mathbf{r}) & = - \text{NICS}_{\alpha\beta}(\mathbf{r}) 
\end{align}

where $\alpha$ and $\beta$ denote tensor component indices $x,y,z$. If we set 
$\mathbf{B}_{ext}=\hat{e}_z=(0,0,1)^T$, {\em i.e.} parallel to the unit vector in $z$ direction (=$\hat{e}_z$) the tensor equation contracts to a vector equation

\begin{align}
   \mathbf{B}_{ind}(\mathbf{r})  & = - \begin{pmatrix}  \text{NICS}_{xz}(\mathbf{r})\\  \text{NICS}_{yz}(\mathbf{r}) \\  \text{NICS}_{zz}(\mathbf{r}) \end{pmatrix}.
\end{align}

The notation can be simplified by defining $\text{NICS}_{xz}=\text{NICS}_{x}$, and so on to

\begin{align}
   \mathbf{B}_{ind}(\mathbf{r})  & = - \begin{pmatrix}  \text{NICS}_{x}(\mathbf{r})\\  \text{NICS}_{y}(\mathbf{r}) \\  \text{NICS}_{z}(\mathbf{r}) \end{pmatrix}.
\end{align}

Since both, the external field $\mathbf{B}_{ext}$ and also the total field $\mathbf{B}_{tot}$ are subject to classical electromagnetism (at least in a statistical sense), also the Ampére-Mawell law (here in its differential form) must be fulfiled

\begin{align}
 \nabla\times\mathbf{B} &= \mu_0 \left(\mathbf{J} + \frac{\partial \mathbf{E}}{\partial t}\right).
\end{align}

Then for $\mathbf{B}_{ind}$, where $\mu_0$ is the vacuum permeability, $\mathbf{E}$ an electric field and in the static case $\frac{\partial \mathbf{E}}{\partial t} = 0$ we obtain

\begin{align}
  \mathbf{J} &= \mu_0^{-1} \nabla\times\mathbf{B}_{ind} 
\end{align}

and

\begin{align}
  \mathbf{J} & = -\mu_0^{-1} \nabla\times\begin{pmatrix}  \text{NICS}_{x}(\mathbf{r})\\  \text{NICS}_{y}(\mathbf{r}) \\  \text{NICS}_{z}(\mathbf{r}) \end{pmatrix} \label{main},
\end{align}

here, $\mathbf{J}$ corresponds to the CD which is induced by the the external field $\mathbf{B}_{ext}$ and sometimes is also denoted as $\mathbf{J}^{\mathbf{B}_{ext}}$ or similar and is usually called magnetically induced current density (MICD). Now one is faced with the problem of computing spatial derivatives of the NICS components in order to evaluate the curl operator
in equation (\ref{main}). Unfortunately such derivatives are not available in current computational chemistry codes, hence we need to do numerical approximations. For that we replace the analytical derivation by numerical derivatives. Defining forward differences $\Delta$

\begin{align}
  \Delta V_\alpha(\epsilon) = V_\alpha(x+\delta(x,\epsilon)h,y+\delta(y,\epsilon)h,z+\delta(z,\epsilon)h) -
  V_\alpha(x,y,z)
\end{align}
of a differentiable vector field $\mathbf{V}(x,y,z)=(V_x(x,y,z),V_y(x,y,z),V_z(x,y,z))^T$, finite $h>0$ and the Kronecker delta function $\delta$. With this a simple numerical approximation to the curl of $\mathbf{V}$ like is

\begin{align}
(\nabla \times \mathbf{V})_\alpha \approx \frac{1}{h} \epsilon^{\alpha\beta\gamma} \Delta V_\gamma(\beta) \label{numcurl}
\end{align} 

where we use the Levi-Civita symbol $\epsilon^{\alpha\beta\gamma}$ and the Einstein tensor summation convention. By inserting $\mathbf{J}$ from (\ref{main}) in (\ref{numcurl}) the numerical approximation $\tilde{\mathbf{J}}$ to $\mathbf{J}$ is yielded

\begin{align}
J_\alpha \approx \tilde{J}_\alpha = - \frac{1}{\mu_0 h} \epsilon^{\alpha\beta\gamma} \Delta \text{NICS}_\gamma(\beta) \label{numcurl_num}
\end{align} 

which together with (\ref{main}) represents the main result of this work.

An interesting property of the currents \( \tilde{\mathbf{J}} = (\tilde{J}_x, \tilde{J}_y, \tilde{J}_z)^T \), as compared to other numerical approximations to \( \mathbf{J} \) obtained from standard quantum chemical software~\cite{Berger1}, is that the analytical and defining property of \( \mathbf{J} \)—namely that it is divergence-free:

\begin{align}
\nabla \cdot \mathbf{J} = 0,
\end{align}

for any stationary eigenstates of the system—still holds for \( \tilde{\mathbf{J}} \) to high accuracy, provided that the numerical representation of the NICS field is sufficiently smooth. This is particularly noteworthy because non-zero divergences can pose significant challenges for topological analyses, where substantial deviations from zero-divergence often arise in calculations based on incomplete basis sets or perturbative expansions of the wave function.

At this point it becomes apparent that $\tilde{\mathbf{J}}$ is not equal to the CD obtained from numerical approximations to the magnetically perturbed wavefunction and the original defintion by Schrödinger. Monaco, Summa, Zanasi and one of us have elaborated on this subject in detail in ref.\cite{Monaco2024}. In summary from this perspective, such non divergence-free numerical approximations $\overset{\approx}{\mathbf{J}}$ to $\mathbf{J}$ contain in contrast to $\tilde{\mathbf{J}}$ a spurious contamination which essentially can be described as the gradient of the Possion potential of the spurious non-zero divergence,

\begin{align}
\phi_{spurious} = -\frac{1}{4\pi}\int \frac{\mathbf{r}-\mathbf{r}'}{|\mathbf{r}-\mathbf{r}'|^3} \nabla'\cdot\overset{\approx}{\mathbf{J}}(\mathbf{r}')d^3\mathbf{r}' \label{spurious}
\end{align}
subtraction of the term then yields $\tilde{\mathbf{J}}$\cite{Monaco2024}
\begin{align}
 \tilde{\mathbf{J}} = \overset{\approx}{\mathbf{J}} -\nabla\phi_{spurious},
\end{align}

where now $\nabla \cdot \tilde{\mathbf{J}} = 0$. A handful examplary calculations of $\tilde{\mathbf{J}}$ are reported and discussed in detail in ref.\cite{Monaco2024}, so we will show only one examplary calculation on the benzene molecule and using Turbomole\cite{TM} for the magntetic response and NICS calculations in the following.

\begin{figure}[htbp]
    \centering

    \begin{subfigure}[b]{0.49\textwidth}
        \centering
        \includegraphics[width=1.3\linewidth]{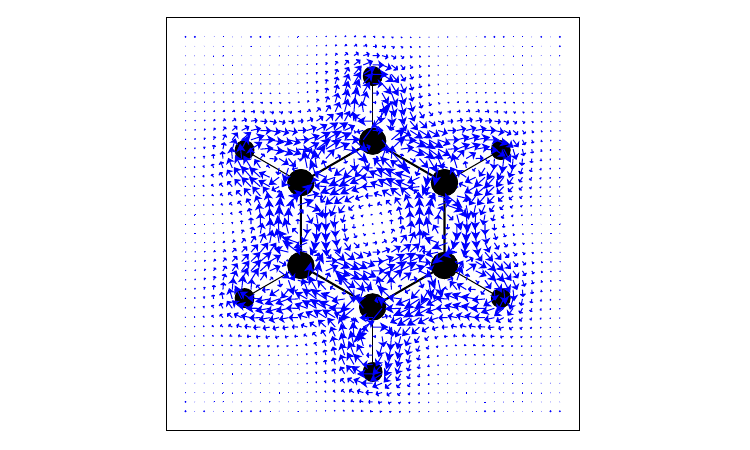}
        \caption{$\tilde{\mathbf{J}}$ computed via (\ref{numcurl_num}).}
        \label{fig:subfig1}
    \end{subfigure}%
    \begin{subfigure}[b]{0.49\textwidth}
        \centering
        \includegraphics[width=1.3\linewidth]{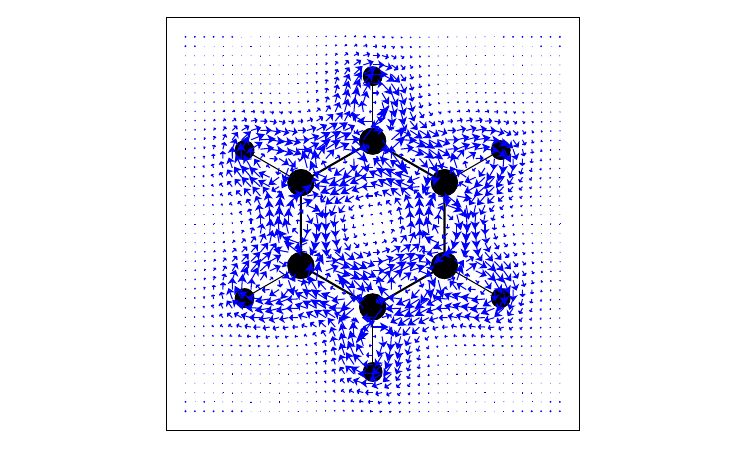}
        \caption{$\mathbf{J}^\text{GIMIC}$ computed with GIMIC\cite{gimic1,gimic2}.}
        \label{fig:subfig2}
    \end{subfigure}
    \vskip\baselineskip

    \begin{subfigure}[b]{0.48\textwidth}
        \centering
        \includegraphics[width=1.3\linewidth]{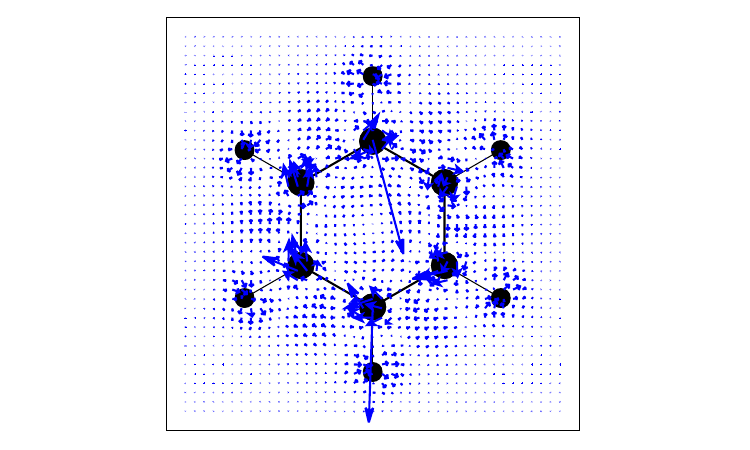}
        \caption{$\tilde{\mathbf{J}}-\mathbf{J}^\text{GIMIC}$ scaled by 10.0.}
        \label{fig:subfig3}
    \end{subfigure}
    \begin{subfigure}[b]{0.45\textwidth}
        \centering
        \includegraphics[width=0.85\linewidth]{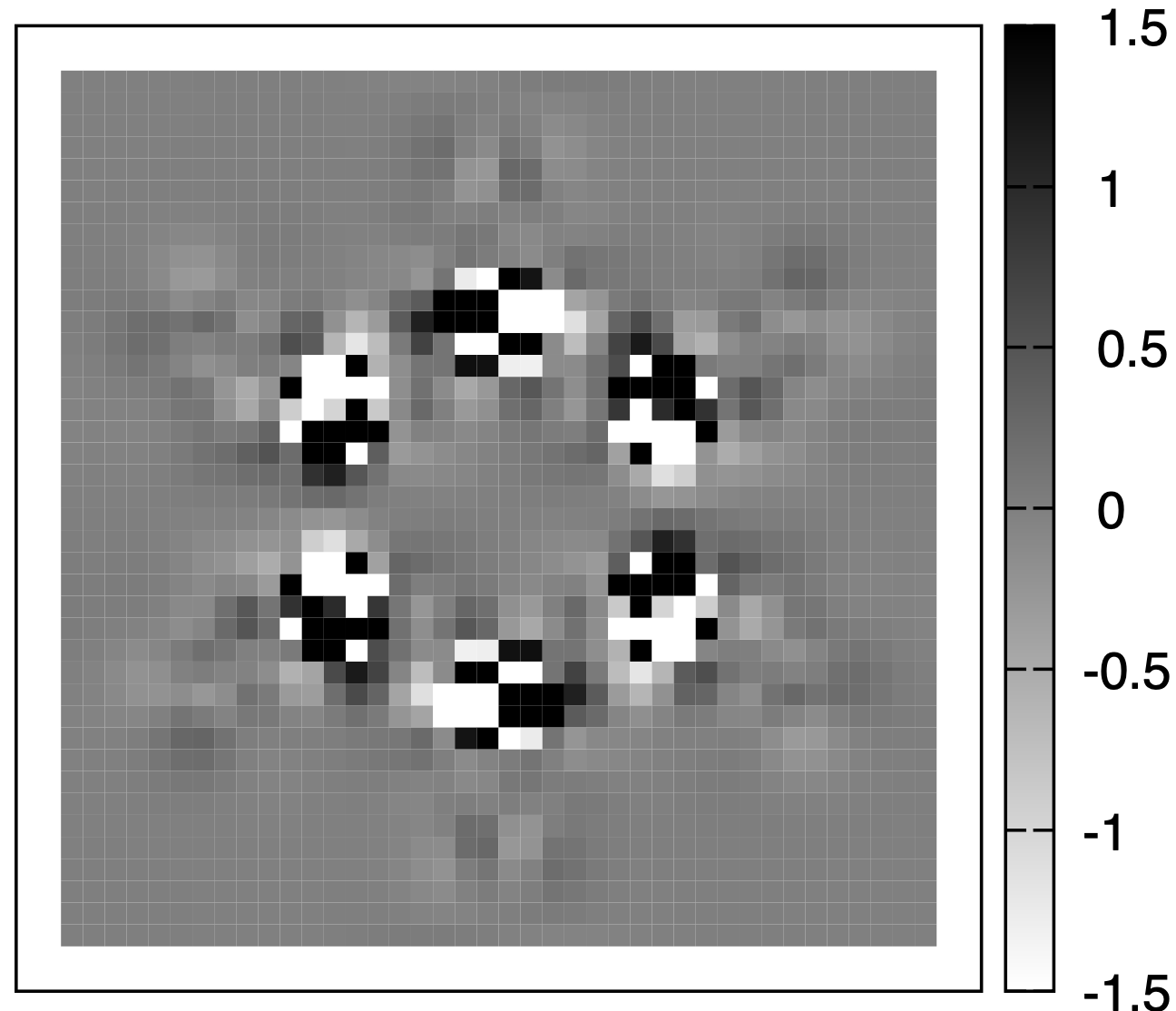}
        \caption{Divergence of $\mathbf{J}^\text{GIMIC}$ ($\nabla\cdot\mathbf{J}^\text{GIMIC}$).}
        \label{fig:subfig4}
    \end{subfigure}

    \caption{CDs and related quantities in the molecular plane of a benzene molecule (C$_6$H$_6$) on a square grid with steps of 0.3 bohr.}
    \label{fig:bigfigure}
\end{figure}

Based on a DFT(PBE0)/def2-SV(P)\cite{PBE0,TM-I,TM-II} optimized structure, nucleus-independent chemical shieldings were computed on a grid at positions based on equation (\ref{numcurl_num}) generated by a python script\footnote{The script can be obtained via request from the authors.}. Here, the grid was chosen to lie in the molecular plane. A plot of the obtained $\tilde{\mathbf{J}}$ vectors is shown in Fig. \ref{fig:bigfigure} along with a plot of the CD obtained with GIMIC\cite{gimic1,gimic2} ($\mathbf{J}^\text{GIMIC}$), the respective differences between the two methods $\tilde{\mathbf{J}}-\mathbf{J}^\text{GIMIC}$ and the divergence of the CD obtained from GIMIC at the same level of theory.

A prima vista the differences between $\tilde{\mathbf{J}}$ (Fig. \ref{fig:subfig1}) and $\mathbf{J}^\text{GIMIC}$ (Fig. \ref{fig:subfig1}) in direct comparison seem only minor, however upon subtracting the one from the other (and upscaling) the differencs become more apparent (Fig. \ref{fig:subfig3}). In particular closer to the nuclei appear lager deviations. The reson for that is the strong correlation of the differences (explicitely given in eq. (\ref{spurious})) with the divergence of $\mathbf{J}^\text{GIMIC}$. It is well known that such divergences accumulate close to the nuclei where basis set incompleteness is more pronounced,\cite{gimic1} which is confirmed by the distribution of $\nabla\cdot\mathbf{J}^\text{GIMIC}$ shown in Fig. \ref{fig:subfig4} 

\section{Conclusion}
We have divised a new scheme to obtain numerical appoximations to the quantum mechanical current density that unlike previously described methods does not directly arise from the (perturbed) wave functions but rather from the chemical shift tensor and its spatial derivatives. This approximate CD is virtually divergence free and can be very simply implemented in any programm that can compute chemical shieldings even by means of simply interfacing script routines.

We currently investigate methods to decompose $\mathbf{J}$ into components for a simplified analyses which are based on the here propsed method.

\section{Acknowledgement}
RB gratefully acknowlegdes many insightful discussions with Prof. Dr. G. Monaco and Prof.
Dr. R. Zanasi from the University of Salerno.

\bibliography{lib}

\begin{thebibliography}{10}

\bibitem{Schroedinger1926-IV}
E.~Schr\"{o}dinger.
\newblock Quantisierung als eigenwertproblem.
\newblock {\em Annalen der Physik}, 386(18):109–139, January 1926.

\bibitem{RingCurrents}
P.~Lazzeretti.
\newblock Ring currents.
\newblock {\em Progress in Nuclear Magnetic Resonance Spectroscopy},
  36(1):1–88, February 2000.

\bibitem{Berger1}
Dage Sundholm, Heike Fliegl, and Raphael~J.F. Berger.
\newblock Calculations of magnetically induced current densities: theory and
  applications.
\newblock {\em WIREs Computational Molecular Science}, 6(6):639–678, June
  2016.

\bibitem{Berger2}
Dage Sundholm, Maria Dimitrova, and Raphael J.~F. Berger.
\newblock Current density and molecular magnetic properties.
\newblock {\em Chemical Communications}, 57(93):12362–12378, 2021.

\bibitem{Hirschfelder}
Joseph~O. Hirschfelder.
\newblock Quantum mechanical equations of change. i.
\newblock {\em The Journal of Chemical Physics}, 68(11):5151–5162, June 1978.

\bibitem{NICS1}
Paul von~Ragué Schleyer, Christoph Maerker, Alk Dransfeld, Haijun Jiao, and
  Nicolaas J.~R. van Eikema~Hommes.
\newblock Nucleus-independent chemical shifts: A simple and efficient
  aromaticity probe.
\newblock {\em Journal of the American Chemical Society}, 118(26):6317–6318,
  January 1996.

\bibitem{NICS2}
Zhongfang Chen, Chaitanya~S. Wannere, Clémence Corminboeuf, Ralph Puchta, and
  Paul von~Ragué Schleyer.
\newblock Nucleus-independent chemical shifts (nics) as an aromaticity
  criterion.
\newblock {\em Chemical Reviews}, 105(10):3842–3888, September 2005.

\bibitem{Monaco2024}
Guglielmo Monaco, Francesco~F. Summa, Riccardo Zanasi, and Raphael J.~F.
  Berger.
\newblock Calculation of divergenceless magnetically induced current density in
  molecules.
\newblock {\em The Journal of Chemical Physics}, 161(19), November 2024.

\bibitem{TM}
Filipp Furche, Reinhart Ahlrichs, Christof H\"{a}ttig, Wim Klopper, Marek
  Sierka, and Florian Weigend.
\newblock Turbomole.
\newblock {\em WIREs Computational Molecular Science}, 4(2):91–100, July
  2013.

\bibitem{PBE0}
C.~Adamo, M.~Cossi, and V.~Barone.
\newblock An accurate density functional method for the study of magnetic
  properties: the pbe0 model.
\newblock {\em Journal of Molecular Structure: THEOCHEM}, 493(1–3):145–157,
  December 1999.

\bibitem{TM-I}
Reinhart Ahlrichs, Michael B\"{a}r, Marco H\"{a}ser, Hans Horn, and Christoph
  K\"{o}lmel.
\newblock Electronic structure calculations on workstation computers: The
  program system turbomole.
\newblock {\em Chemical Physics Letters}, 162(3):165–169, October 1989.

\bibitem{TM-II}
Marco H\"{a}ser and Reinhart Ahlrichs.
\newblock Improvements on the direct scf method.
\newblock {\em Journal of Computational Chemistry}, 10(1):104–111, January
  1989.

\bibitem{gimic1}
Jonas Jusélius, Dage Sundholm, and J\"{u}rgen Gauss.
\newblock Calculation of current densities using gauge-including atomic
  orbitals.
\newblock {\em The Journal of Chemical Physics}, 121(9):3952–3963, September
  2004.

\bibitem{gimic2}
Heike Fliegl, Stefan Taubert, Olli Lehtonen, and Dage Sundholm.
\newblock The gauge including magnetically induced current method.
\newblock {\em Physical Chemistry Chemical Physics}, 13(46):20500, 2011.

\end{thebibliography}
\end{document}